\documentclass[aps,pra,twocolumn,a4paper]{revtex4}
\pdfoutput=1

\usepackage{amsmath}
\usepackage{graphicx}

\begin{document}

\title{Photostop: Production of zero-velocity molecules\\by photodissociation in a molecular beam}

\author{Alexandre Trottier}
\altaffiliation[Current address: ]{AECL Chalk River Laboratories, Deep River, ON, K0J 1J0, Canada.}
\affiliation{Durham University, Department of Chemistry, South Road, Durham, DH1 3LE, U.K.}
\author{David Carty}
\affiliation{Durham University, Departments of Physics and Chemistry, South Road, Durham, DH1 3LE, U.K.}
\author{Eckart Wrede}
\email[Corresponding author: ]{eckart.wrede@durham.ac.uk}
\affiliation{Durham University, Department of Chemistry, South Road, Durham, DH1 3LE, U.K.}

\date{January 25, 2010}

\keywords{cold molecules, deceleration, trapping, photodissociation}

\begin{abstract}
  We have demonstrated a new, accessible and economical technique, dubbed photostop, for producing high densities of trappable molecules. Direct measurements are presented of NO molecules produced with a narrow velocity distribution centered at zero in the laboratory frame. NO$_2$, initially cooled in a pulsed molecular beam, is photodissociated such that the recoil velocity of the NO photofragments cancels out the velocity of the beam. NO(X$^2\Pi_{3/2}, v=0, J=1.5$) molecules are observed up to 10~mircoseconds after the dissociation event in the probe volume at an estimated density of $10^7$ cm$^{-3}$ per quantum state and at a translational temperature of 1.6~K. Through the choice of suitable precursors, photostop has the potential to extend the list atoms and molecules that can be slowed or trapped. It should be possible to accumulate density in a trap through consecutive loading of multiple pulses.
\end{abstract}

\maketitle

\section{Introduction}

The growth over the last decade in the number of experiments designed to create ground state molecules at cold ($<1$~K) and ultracold ($<1$~mK) temperatures is a direct result of the wide-ranging impact expected from the field. It is anticipated that ensembles of ultracold polar molecules will find many applications due to the possibility that their long-range dipolar interactions can be controlled. Such applications include tests of fundamental physical laws, quantum information processing, quantum simulators and controlled molecular dynamics \cite{carr09}.

The long interrogation times, quantum state selectivity and rich internal structure afforded by trapped ultracold molecules offer inherent advantages for measurements that require ultra-high precision. For example, the determination of the dipole moment of the electron would test for time-reversal symmetry violation \cite{hudson02} and the determination of a time variation of fundamental constants would test for the equivalence principle of general relativity \cite{flambaum07}. Ultracold polar molecules that have been trapped in optical lattices could be made to form an adjustable network of variably-interacting quantum bits (qubits) where it would be possible to engineer fast logic gates for quantum information processing \cite{demille02}. Furthermore, such a network could be adjusted to act as a quantum simulator for any fully quantum many-body system for solving problems in condensed matter physics that are intractable using traditional computational methods and hardware \cite{micheli06}. Molecular collisions and chemical reactions depend sensitively on the mutual interactions between the molecules. At cold and ultracold temperatures, the collision energy is similar to the Stark and Zeeman shifts experienced by the molecules in external fields. Therefore, by applying a tailored external electrical or magnetic field, molecular interactions can be controlled to enhance or suppress specific outcomes of a chemical reaction via the fine-tuning of scattering resonances \cite{krems08}.

To date, the only polar molecules that have been produced in their rovibronic ground states at ultracold temperatures are KRb \cite{ni08}, RbCs \cite{sage05} and LiCs \cite{deiglmayr08}, which were formed by associating ultracold atoms. The technique is thus far limited to alkali-alkali molecules, which do not possess a magnetic moment, a desirable property for many applications. In a promising development, it was shown that it may be possible to laser cool SrF directly to ultracold temperatures due to its unique energy level structure \cite{shuman09}. However, for applications, a variety of species at ultracold temperatures and high densities is required. The most versatile approach is to sympathetically cool trapped \emph{cold} polar molecules into the ultracold regime using ultracold atoms as a refrigerant \cite{wallis09}. Therefore, the initial production of high densities of a variety of cold molecular species remains a major challenge in the field.

A variety of techniques have been developed for the production of cold ground state molecules \cite{narevicius08,patterson09,kay09} with buffer gas cooling and Stark deceleration the established leaders in the field \cite{campbell07,vandemeerakker08}. Each technique has well documented limitations, not least in terms of expense and difficulty of execution. Therefore, it is very important to continue developing new economical techniques that have the potential to expand the list of cold species.

\section{Photostop technique}

\begin{figure}
  \includegraphics[width=8.6cm]{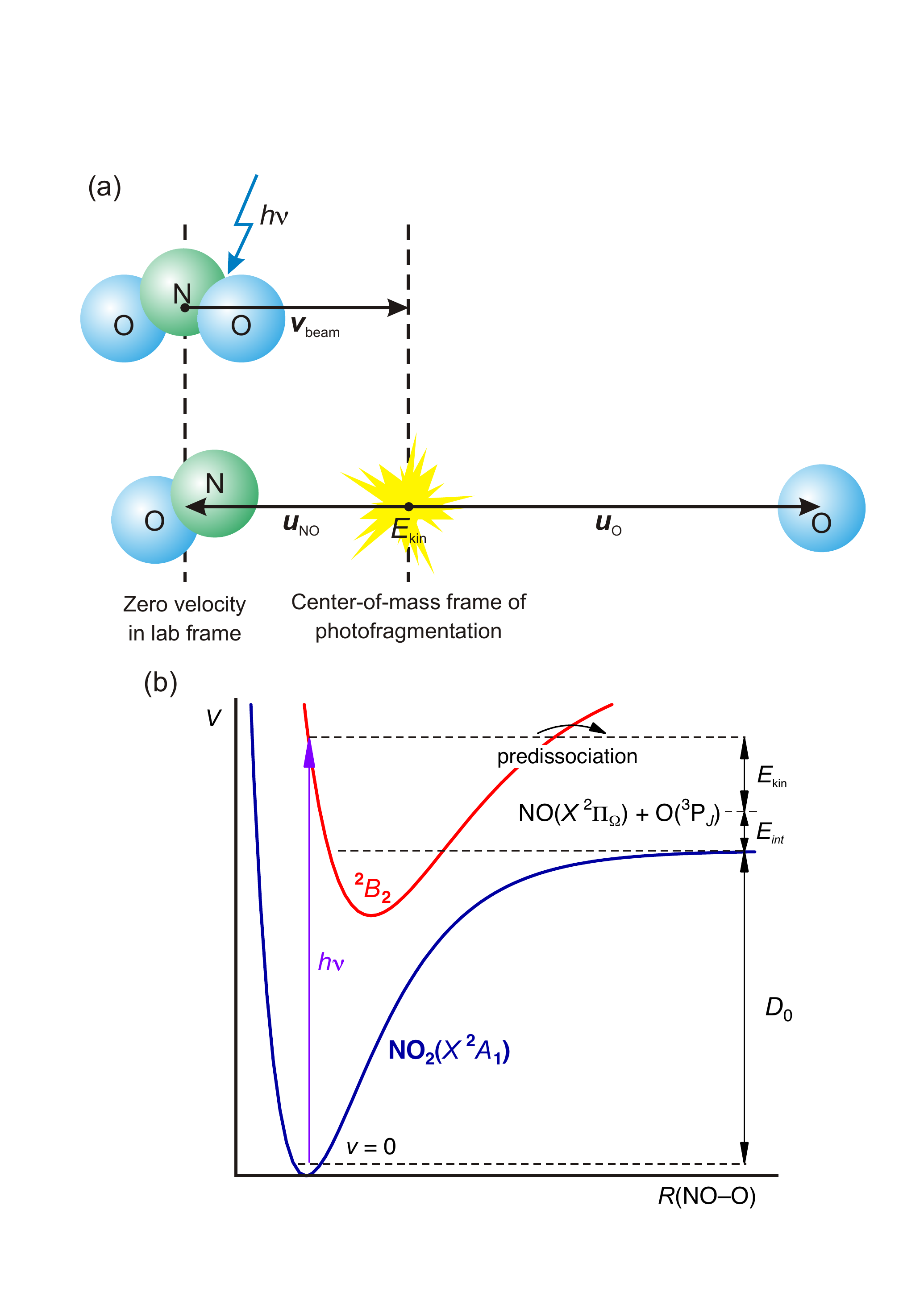}
  \caption{\label{fig1} (Color online)
    (a) Schematic outline of the photostop technique.
    The recoil velocity of NO fragments, $\boldsymbol{u}_\text{NO}$, from the photodissociation of NO$_2$ molecules cancels the velocity of the molecular beam, $\boldsymbol{v}_\text{beam}$.
    Therefore, NO molecules are produced with zero velocity in the laboratory frame while the oxygen co-fragments are accelerated.
    (b) Schematic diagram of the NO$_2$ potential energy surfaces.
    Ground state NO$_2$ molecules are photo excited into the $^2B_2$ state which predissociates onto the ground state surface to form NO($X~^2\Pi_\Omega$) molecules and O($^3P_J$) atoms.
    By tuning the photon energy, $h\nu$, the total kinetic energy, $E_\text{kin}$, released by the photodissociation process can be adjusted for given total internal energy, $E_\text{int}$, of both photofragments.
  }
\end{figure}

In this paper, we present direct measurements of molecules produced with zero mean velocity in the laboratory frame, using a novel, accessible and economical technique, dubbed photostop. The method is based on the controlled breaking of one chemical bond, via photodissociation, in a precursor molecule initially cooled in a supersonic molecular beam.
The result is an atomic or molecular fragment that recoils with a velocity that cancels out the initial velocity of the precursor molecule in the beam, $v_{\text{beam}}$, as illustrated in FIG.~\ref{fig1}(a).

In this experiment the precursor molecule of choice is NO$_2$. As can be seen in FIG.~\ref{fig1}(b), photon energies, $h\nu$, exceeding the dissociation energy, $D_0$, of the NO--O bond cause the NO$_2$ molecule to fragment into an NO molecule and an oxygen atom. The excess energy, $E_\text{exc} = h\nu - D_0$, is partitioned between the kinetic and internal energies of the fragments in varying proportions determined by the complex predissociation dynamics~\cite{Matthews07}.
For fragments formed in a given state, the total internal energy, $E_\text{int}$, is fixed and the total kinetic energy, $E_\text{kin}$, released is therefore governed only by the photon energy
\begin{equation}
  E_\text{kin} = E_\text{kin}(\text{NO}) + E_\text{kin}(\text{O}) = h\nu - D_0 - E_\text{int}, \label{eqn:TKER}
\end{equation}
where $E_\text{kin}(\text{NO})$ and $E_\text{kin}(\text{O})$ are the kinetic energies of the NO and O fragments, respectively.
By conservation of momentum, the velocities of the fragments (in the center-of-mass frame of the photodissociation process) are given by
\begin{subequations}\label{eqn:ufrag}
\begin{eqnarray}
  u_\text{NO} &=& \sqrt{2\,E_\text{kin}\cdot\frac{m_\text{O}}{m_\text{NO}\,m_{\text{NO}_2}}} \\
  u_\text{O} &=& \sqrt{2\,E_\text{kin}\cdot\frac{m_\text{NO}}{m_\text{O}\,m_{\text{NO}_2}}}
\end{eqnarray}
\end{subequations}
where $m_\text{O}, m_\text{NO}$ and $m_{\text{NO}_2}$ are the respective masses.
NO molecules are stopped in the laboratory frame if $\boldsymbol{u}_\text{NO}=-\boldsymbol{v}_\text{beam}$ and, from equations~\ref{eqn:TKER} and~\ref{eqn:ufrag}, the wavelength of the photons required is given by
\begin{equation}
  \lambda_\text{NO}=hc\left[\frac{1}{2}\left(\frac{m_\text{NO}m_{\text{NO}_2}}{m_\text{O}}\right)v_{\text{beam}}^2+D_0+E_{\text{int}}\right]^{-1} \label{eqn:lambda}
\end{equation}
Exchanging the indices of NO and O in Eq.\ \ref{eqn:lambda} yields the corresponding wavelength to stop the O fragment.

\section{Experiment}

\begin{figure*}
  \includegraphics[width=11.5cm]{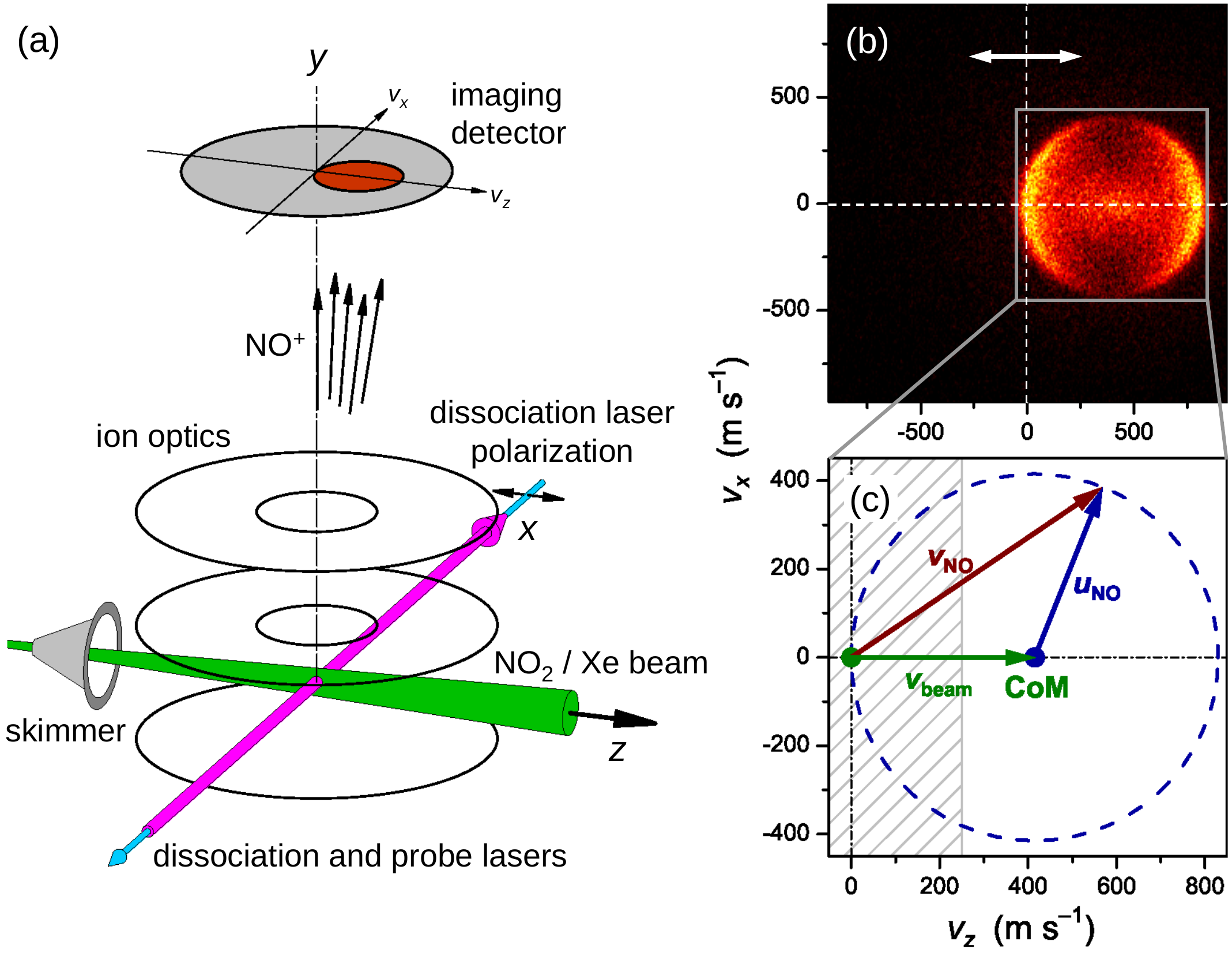}
  \caption{\label{fig2} (Color online)
    (a) Schematic diagram of the experiment.
    A pulsed molecular beam of NO$_2$ seeded in Xe is skimmed and intersected at right angles by the photodissociation and probe laser beams.
    The ionized NO photofragments are accelerated by static electric fields in a velocity map configuration (ion optics) towards a position sensitive detector to image the ions.
    (b) Velocity mapped ion image of NO($X\,^2\Pi_{3/2}, v=0, J=1.5$) fragments from the photodissociation of NO$_2$ at 386.4~nm.
    The image shows the distribution of NO lab velocities projected onto the plane of the molecular and laser beams.
    The lab velocity origin is indicated by the dotted lines and the arrow indicates the axis of cylindrical symmetry of the dissociation process and the polarisation direction of the dissociation laser.
    (c) Newton diagram of the photodissociation process in the $(v_z,v_x)$-plane of the experiment.
    The recoil velocity of the NO fragment, $\boldsymbol{u}_\text{NO}$, in the center-of-mass (CoM) frame can lie at any point on the Newton sphere, indicated by the dashed circle, weighted by the anisotropy of the photodissociation process.
    The NO velocity in the laboratory frame, $\boldsymbol{v}_\text{NO}$ is the vector sum of the molecular beam velocity, $\boldsymbol{v}_\text{beam}$, and $\boldsymbol{u}_\text{NO}$.
    The necessary kinematic conditions for photostop are met when the Newton sphere intersects the velocity origin.
    The grey shaded area represents the $v_z$-velocity acceptance of the probe volume after 1~$\mu$s time delay (\emph{c.f.}\ FIG.~\ref{fig3}(a)).
  }
\end{figure*}

A schematic diagram of the experiment is shown in FIG.~\ref{fig2}(a).
The molecular beam is formed from a skimmed pulsed (10~Hz repetition rate) supersonic expansion of 1--5\% NO$_2$  in 4~ bar of Xe, the direction of which defines the $z$-axis of the experiment.
The molecular beam is intersected at right angles by two laser beams counter-propagating along the $x$-axis.
The first beam is used to photodissociate the NO$_2$ molecules and is generated by a pulsed dye laser tunable in the range of 380--400~nm.
The second beam, generated by another pulsed dye laser tunable around 226~nm, is used to ionize NO($X\,^2\Pi$) molecules in a specific rovibrational quantum state using (1~+~1) resonance-enhanced multiphoton ionization (REMPI) via the $A\,^2\Sigma$ state.
The molecular beam and the laser beams intersect at the center of an ion lens system oriented along the $y$-axis.
The resulting NO$^+$ ions are accelerated by the electrostatic field towards a position sensitive imaging detector consisting of a pair of micro-channel plates, a phosphorescent screen and a CCD camera.
The ion lens system is designed and operated such that the velocity of the original NO molecule is mapped onto a unique position on the detector, a technique known as velocity mapped ion imaging (VMI) \cite{Eppink97}.
The detected ion image corresponds to the 2-dimensional projection of the original 3-dimensional velocity distribution of the NO molecules onto the $(v_x,v_z)$ plane.
Such a set-up allows the direct measurement of the absolute velocity of both the molecular beam and of the NO fragment molecules, which is vital in maximizing the efficiency of the technique.

In order to measure the velocity of the NO molecules in the laboratory frame, the velocity origin, \emph{i.e.}\ the position on the detector that corresponds to $(v_x,v_z)~/~\mathrm{m~s^{-1}}=(0,0)$, was determined from the center of an ion image of thermal NO bled into the vacuum chamber.
The velocity of the molecular beam was determined from an ion image of the small fraction of ``native'' NO present in the beam due to the thermal decomposition of the NO$_2$ sample at room temperature.
The velocity of the molecular beam was measured to be $v_\text{beam}=415\text{ m s}^{-1}$ with a full-width at half maximum of 80~m~s$^{-1}$, which corresponds to a translational temperature of the NO$_2$ molecules of 6~K.

There are three sources of NO molecules that can contribute towards signal in the ion image:
native NO, present as a background component in the molecular beam;
thermal background NO that is not pumped away between gas pulses;
and NO produced as a product of the photodissociation.
The ratio of $^2\Pi_{3/2}$ spin-orbit excited to $^2\Pi_{1/2}$ ground state NO molecules present in the beam was measured to be less than 1~:~40, therefore spin-orbit excited NO originates almost exclusively from the NO$_2$ dissociation.
Thus, by aiming to photostop spin-orbit excited NO, interference from native NO present in the molecular beam is minimized.
All ion images were recorded with an automatic shot-by-shot background subtraction by electronically blocking every second shot from the photodissociation laser.
Therefore, any remaining signal from NO present in the background gas in the vacuum chamber is minimized.

FIG.~\ref{fig2}(b) shows an ion image representing the velocity distribution of NO molecules in the $X\,^2\Pi_{3/2}, v=0, J=1.5$ state.
The center of the distribution is offset by the velocity of the molecular beam from the laboratory frame velocity origin, the position of which is illustrated by the cross-hairs.
As shown in FIG.~\ref{fig2}(c), the recoil velocity of the NO fragments, $\boldsymbol{u}_\text{NO}$, can lie on any point on the Newton sphere.
However, the photodissociation of NO$_2$ is strongly anisotropic with the NO fragments preferentially recoiling in either direction along the polarization direction of dissociation light, as can be seen in FIG.~\ref{fig2}(b).
The outer ring in the image corresponds to the recoiling NO fragments, where the polarization direction has been aligned along the molecular beam ($z$) axis.
With the dissociation laser wavelength tuned to 386.4~nm, the peak of the NO density was optimally overlapped with the velocity origin.
Those NO molecules coinciding with the velocity origin are thus photo-stopped and those that appear on the exact opposite side of the ion image are accelerated to twice the molecular beam speed.
According to Eq.\ 3, with $D_0=25128.6$~cm$^{-1}$ \cite{Jost96}, 386.4~nm corresponds to a NO fragment speed of 418.5~m s$^{-1}$, which equals the molecular beam speed of 415~m s$^{-1}$ to within the nearest pixel on the camera.

\section{Stopped molecules}

\begin{figure*}
  \includegraphics[width=11.5cm]{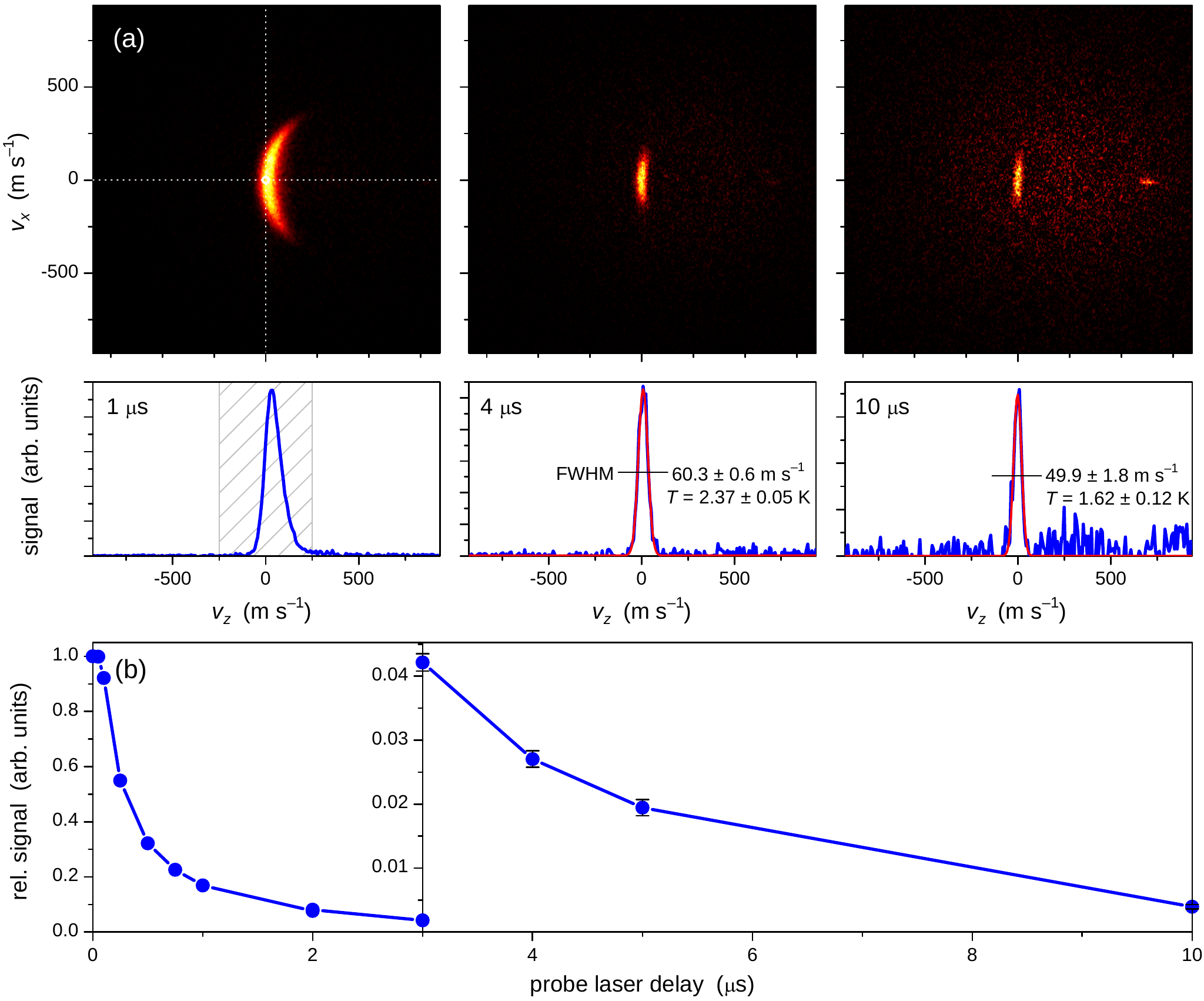}
  \caption{\label{fig3} (Color online)
    (a) Velocity mapped ion images of NO($X~^2\Pi_{3/2}, v=0, J=1.5$) fragments from the photodissociation of NO$_2$ under photostop conditions at 386.4~nm and at different time delays between the dissociation and probe lasers.
    The velocity distribution projected onto the molecular beam axis ($v_z$) is shown underneath each ion image.
    The grey shaded area represents the $v_z$-velocity acceptance of the probe volume after 1~$\mu$s time delay.
    One-dimensional Maxwell-Boltzmann velocity distributions are fitted to the profiles at 4 and 10~$\mu$s (red, online).
    The full-widths-at-half-maximum (FWHM) and the corresponding translational temperatures obtained from the fit are indicated.
    (b) Decay of the integrated NO signal for each recorded ion image as a function of time delay.
  }
\end{figure*}

To demonstrate that NO fragments are indeed stopped in the laboratory frame, we have recorded a series of ion images, shown in FIG.~\ref{fig3}(a), as a function of the time delay between the 5~ns long pulses of the photodissociation and probe lasers.
It is clear that molecules remain in the probe laser volume even after 10~$\mu$s.
The panels below  each of the ion images show the projection of that ion image onto the $v_z$ axis.
After 1~$\mu$s delay, it can be seen that those NO fragments that recoil in the same direction as the molecular beam (positive $v_z$) have duely left the probe laser volume.
The probe volume is much larger along the $x$ axis than in the other dimensions, therefore, only NO molecules with small $v_y$ and $v_z$ components are discriminated by the probe laser.
On the contrary, molecules with larger $v_x$ components are detected.
The grey shaded areas in Fig.~2C and Fig.~3A represent the velocity acceptance of the probe volume after 1~$\mu$s delay.
All recoiling NO molecules with velocities that lie within this velocity acceptance appear as a crescent shape in the 1~$\mu$s ion image.
The asymmetry on the positive side of the corresponding $v_z$ profile is simply due to the projection of the crescent onto the $v_z$ axis.
After a delay of 4~$\mu$s, and subsequently 10~$\mu$s, the ion images appear straighter because the velocity acceptance narrows in both $v_y$ and $v_z$ but not in $v_x$.
The corresponding projections are almost symmetrical, thus, a temperature of $T=2.37 \pm 0.05$~K and $1.62 \pm 0.12$~K can be assigned to the NO molecules that remain at 4 and 10~$\mu$s, respectively, by fitting the profiles with a one-dimensional Maxwell-Boltzmann distribution of the form
\begin{equation}\label{eq:Boltzmann}
  f(v_z) \propto \exp\left(-\frac{m_\text{NO} v_z^2}{2kT}\right).
\end{equation}

FIG.~\ref{fig3}(b) shows the decay of the normalized integrated NO signal in each recorded ion image as a function of delay time.
The signal decays rapidly in the first microsecond due to NO molecules that recoil in the direction of the molecular beam leaving the probe volume.
This is followed by a slow decay that represents the remaining slow molecules leaving the probe volume.
After 10~$\mu$s, $0.39 \pm 0.05$\% of NO($X\,^2\Pi_{3/2}, v=0, J=1.5$) fragments present at zero delay remain.

Clearly, a large proportion of the NO fragments produced are not photo-stopped.
The velocities of most NO fragments do not lie within the velocity acceptance of the probe volume and are therefore lost at later delays.
This is a direct result of the angular distribution of the photodissociation process, as can be seen in FIG.~\ref{fig2}(c).
The velocity spread of the molecular beam reduces further the proportion of NO fragments that can be photo-stopped.
At a delay of 10~$\mu$s, the $v_z$ velocity spread of the detected NO fragments is $\pm 25$~m~s$^{-1}$ (see FIG.~\ref{fig3}(a)) compared to the $\pm 40$~m~s$^{-1}$ velocity spread of the molecular beam.
However, the reduction of signal by a factor of approx.\ 250 after this time is mainly accounted for by the unfavorable angular distribution.

In these experiments, only NO fragments in the $X\,^2\Pi_{3/2}, v=0, J=1.5$ state were detected.
However, the photon energy used is 751~cm$^{-1}$ above the dissociation threshold, thus, higher NO rotational levels up to $J=20.5$ and $J=18.5$ in the $^2\Pi_{1/2}$ and $^2\Pi_{3/2}$ spin-orbit states, respectively, are observed to be populated by the photodissociation process.
Although energetically allowed, no evidence of spin-orbit excited O($^3P_{1,0}$) co-fragments was found.
In order to determine which proportion of NO fragments is produced in the $X\,^2\Pi_{3/2}, v=0, J=1.5$ state by the photodissociation process, the state distribution needs to be known.
Unfortunately, the state distribution is as yet unknown at the photodissociation energy of interest and cannot be determined with confidence in our experiments due to the non-linear REMPI detection scheme employed.
However, assuming a statistical state distribution we estimate the proportion to be on the order of 1\%.
Assuming a NO$_2$ density in the probe volume of $10^{12}$~cm$^{-3}$, and taking into account the losses due to the state distribution and the velocity acceptance of the probe volume, we estimate the density of photo-stopped NO molecules that are probed at 10~$\mu$s delay to be $10^7$~cm$^{-3}$.
The distance between the nozzle and the laser beams is large at 150~mm, which is due to the fact that the experimental set-up is not specifically designed for photostop.
In a custom designed machine this distance can feasibly be reduced to 50~mm, thus increasing the NO$_2$ density by an order of magnitude, and therefore increasing the density of photo-stopped NO molecules to $10^8$~cm$^{-3}$.

The detected NO molecules are not the only NO fragments that are photo-stopped.
From a combination of the velocity spread of the molecular beam and the velocity acceptance of the probe volume we estimate that NO molecules in the quantum states $^2\Pi_{1/2}, J=0.5-12.5$ and $^2\Pi_{3/2}, J=1.5-9.5$ are present in the probe volume after 10~$\mu$s in varying proportions.
The density of \emph{all}\/ photo-stopped molecules (ca.\ 10\% of the state distribution) in a custom designed machine is therefore estimated to be on the order of $10^9$~cm$^{-3}$.

During the course of this work, we became aware that Zhao \emph{et al.}\ were independently utilizing the same idea, \emph{i.e.}\ to slow molecular fragments via the recoil from a photodissociation. They recently reported that they had slowed NO($^2\Pi_{1/2}, v=1, J\sim 8.5$) molecules to speeds centered around 130~m~s$^{-1}$ along the molecular beam axis following the 355~nm dissociation of NO$_2$ \cite{zhao09}.
Assuming a strictly monoenergetic molecular beam, they estimated that 3--4\% of all probed NO fragments were in the slowest velocity interval of their distribution between 35 and 45~m~s$^{-1}$.
Crucially, their experimental set-up did not allow them to measure the velocity component along the molecular beam axis directly, but instead they relied on numerical simulations to infer the velocities they quote.
However, they did measure that NO molecules of indeterminate velocity were probed at times up to 300~ns after dissociation.

\section{Prospects}

The implementation of the photostop technique is relatively simple as it is based on well established knowledge and technology.
The technique is economical in its requirements: a molecular beam source; a tunable photodissociation laser; and a suitable laser spectroscopic detection method, \emph{e.g.}\ laser induced fluorescence.
The velocity map ion imaging detection technique is a very effective analytical tool for the success of these experiments because it allows the direct measurement of the molecular beam and photofragment velocities.

We are currently implementing a magnetostatic trap constructed from permanent magnets that will confine photo-stopped NO($^2\Pi_{3/2}, v=0, J=1.5$) molecules in the low field seeking $M_J=3/2$ state at a temperature on the order of 1~K.
The velocity acceptance of such a trap will be comparable to that of the probe volume at 10~$\mu$s delay (see Fig.\ 3A).

In contrast to other techniques that have been used to trap molecules \cite{vandemeerakker08}, a major advantage of the photostop technique is that any trap employed does not have to be opened to allow molecules to enter as the trappable molecules would be produced \emph{in situ}.
Multiple pulses of stopped molecules could therefore be accumulated to increase density in the trap although trap losses due to collisions with the ``unused'' parts of subsequent gas pulses will need to be minimized, for example, by using beam chopping techniques.

Out of the several groups that have proposed schemes for accumulating molecules in a trap, so far, only two have demonstrated this experimentally.
Campbell \emph{et al.}\ have used buffer gas cooling within a superconducting magnetostatic trap to accumulate ground state NH molecules, although the trap lifetime is limited by the continued presence of the buffer gas \cite{campbell07} and Heiner \emph{et al.}\ have loaded two pulses into a molecular synchrotron \cite{heiner07}.
Meijer and co-workers are attempting to accumulate ground state NH molecules in a magnetostatic trap using a novel laser excitation scheme \cite{vandemeerakker06}.
DeMille \emph{et al.}\ have proposed that similar laser excitation schemes could be employed together with a buffer gas molecular beam technique to accumulate molecules in a microwave trap \cite{demille04}.
All of these techniques are technically demanding in comparison to the economy of the photostop technique.

A suitable precursor for the photostop technique is a molecule that absorbs at wavelengths that satisfy Eq.~\ref{eqn:lambda} for typical molecular beam speeds.
While NO$_2$ clearly satisfies this criterion, the bent nature of its $^2A_1$ ground and $^2B_2$ excited states favors rotational excitation of the NO fragment.
This leads to a broad state distribution that dilutes the number of photo-stopped molecules and reduces the state selectivity of the technique.
As mentioned earlier, oxygen atoms can also be stopped by the photodissociation of NO$_2$.
However, by the same argument, NO$_2$ is not the most effective precursor.
Polyatomic precursor molecules that are close to linear in both the ground and excited states are more favorable due to the small likelihood that the excess energy in the dissociation will be partitioned into rotational excitation of the molecular fragments.
They are even more effective if they can be dissociated below the thresholds for vibrational excitation of the fragments.
Effective precursors for producing photo-stopped atoms are likely to be diatomic molecules as the number of available electronic states is small given typical excess energies required for photostop.

\section{Conclusions}
We have demonstrated a novel technique for stopping molecules.
In these experiments we have shown, via the photodissociation of NO$_2$ moving in a molecular beam, that NO fragments can be brought to a standstill in the laboratory frame and observed up to 10~$\mu$s at a density of \emph{ca.}\ $10^7$~cm$^{-3}$.
With the right choice of precursor molecule, we believe that the photostop technique complements existing techniques for slowing molecules and atoms and can add to the growing list of species that can be studied.
The technique is particularly suited towards trapping of molecules or atoms and offers the opportunity to accumulate them in a trap.
The experimental equipment and techniques employed are economical and well enough studied to make them accessible to many laboratories around the world that wish to perform experiments using slow molecules.


\begin{acknowledgements}
We would like to thank the EPSRC for funding (GR/S22783/01 and EP/D055237/1). We also thank Miss Laura Harris, Mr Rob Rae and Mr Oliver Willis for their help. We also acknowledge Prof.\ Tim Softley for useful discussions and Prof.\ Jeremy Hutson for his scrutiny of the manuscript.
\end{acknowledgements}
\clearpage

\end{document}